\documentstyle[12pt]{article}
\newcommand{\be}{\begin{equation}}

\newcommand{\ee}{\end{equation}}

\input tcilatex
\QQQ{Language}{
American English
}

\begin{document}

\begin{titlepage}
%\begin{flushright}
%{\em Draft version}\\
%To be presented for publication in\\ {\em Lett. Math. Phys.}
%\end{flushright}

\vspace{2.5cm}
\begin{center}{\large\bf  Hopf Structure in   Nambu-Lie  $n$-Algebras}\\
 \vspace{1cm}
 A. E. SANTANA  \footnotemark
 \footnotetext{ ademir@ufba.br}
 and R. MURADIAN \footnotemark
 \footnotetext{ muradian@ufba.br}

\vspace{0.5cm}
Instituto de F\'{\i}sica \\
Universidade Federal da Bahia \\
Campus da Ondina \\
40210-340, Salvador, Bahia, Brasil \\

\vspace{0.5cm}
April, 1997
\vspace{3cm}

\begin{abstract}
We give a definition and study Hopf structures
 in ternary ( and $ n $-ary) Nambu-Lie algebra. The fundamental concepts   of
3-coalgebra, 3-bialgebra and Hopf 3- algebra are introduced.
 Some examples of  Hopf structures are   analyzed.

\vspace{0.5cm}


\end{abstract}

\end{center}
\end{titlepage}
%\section{Introduction}

In 1973 Yoishira Nambu \cite{nambu} proposed a generalization of classical
Hamiltonian mechanics, using ternary and higher-order brackets ( $n$-ary
brackets or multibrackets). During the past two decades Nambu proposal has
been a matter for many investigations \cite{flato1} - \cite{gau} and the
permanent interest for this issue is related with the recognition of the
feasible physical richness and the mathematical beauty of ternary and higher
algebraic systems. Recently such an algebraic structure has been analyzed
and reformulated by Tachtajan \cite{tacht1} -- \cite{tacht3} in an invariant
geometrical form. He proposed the notion of Nambu-Lie ''gebra'', which is a
generalization of Lie algebras for ternary (in general $n$-ary) case. The
ternary algebra or ''gebra'' is a linear space in which conditions of
antisymmetry and generalized Jacobi identity are fulfilled.

In this letter we investigate some new aspects of the Nambu proposal,
connected with Hopf algebra concept. That is, a ternary (and $n$-ary) Hopf
structure is introduced as generalization of the usual Hopf algebra, and
some examples are presented.

Let us begin with a definition of a ternary (associative) algebra.

\begin{definition}
. A ternary algebra with unit over a commutative ring {\bf C} is a vector
space $A$ together with a way of multiplying three elements $a,b,c\in A$ 
\begin{equation}
m:A\otimes A\otimes A=A^{\otimes 3}\rightarrow
A,\;\;\;such\;\;that\;\;\;m(a\otimes b\otimes c)=abc
\end{equation}
The unit element in A is thus defined: 
\begin{equation}
m(1\otimes 1\otimes a)=m(1\otimes a\otimes 1)=m(a\otimes 1\otimes 1)=a;
\end{equation}
and the 3-associativity means 
\begin{equation}
(abc)de=a(bcd)e=ab(cde).
\end{equation}
\end{definition}

\begin{definition}
. A 3- associator can be defined by

\begin{equation}
I^{(3)}=2(abc)de-a(bcd)e-ab(cde),
\end{equation}
\end{definition}

Let us give examples of an associative and non-associative 3-algebras.

\begin{example}
{\sc . }{\em (Associative 3-algebra.) } Let $A=\{a,b,c...;m\}$ be a set of
linear operators on a Hilbert space, such that $m$ is the usual associative
product, then $I^{(3)}=0$.
\end{example}

\begin{example}
. {\em (Non-associative 3-algebra. )} Consider $A=\{a,b,c...;m\}$ a set of
analytic functions defined on ${\cal R}^3$, such that 
\[
m(a\otimes b\otimes c){({\bf x)}}=\frac{\partial a{({\bf x)}}}{\partial x_1}%
\frac{\partial b{({\bf x)}}}{\partial x_2}\frac{\partial c{({\bf x)}}}{%
\partial x_3},
\]
where ${\bf x}=(x_1,x_2,x_3)$. In this case, $I^{(3)}\ne 0$. Indeed,
denoting 
\[
\partial _i=\frac \partial {\partial x_i},\,\,\,\,\,\,\,\,\,\partial _{ij}=%
\frac{\partial ^2}{\partial x_i\partial x_j},\,\,\,\,\,\,\,\,\,\,\,%
\,i,j=1,2,3,
\]
$\,\,\,$we obtain the following non zero value for the 3-associator

\begin{eqnarray*}
I^{(3)} &=&2(\partial _1a\,\,\partial _{12}b+\partial _{11}a\,\,\partial
_2b)\;\partial _3c\,\,\partial _2d\;\partial _3e \\
&&\ +\partial _1a\;\partial _2b\;\partial _{13}c\;\partial _2d\;\partial _3e
\\
&&\ -\partial _1a\;\partial _2b\;\partial _1c\;\partial _{23}d\;\partial _3e
\\
&&\ -\partial _1a\;\partial _1b\;\partial _2c\;\partial _{23}d\;\partial _3e
\\
&&\ -\partial _1a\;\partial _{12}b\;\partial _2c\;\partial _3d\;\partial _3e
\\
&&\ -\partial _1a\;\partial _1b\;\partial _{22}c\;\partial _3d\;\partial _3e
\\
&&\ -\partial _1a\;\partial _2b\;\partial _1c\;\partial _2d\;\partial _{33}e.
\end{eqnarray*}
\end{example}

\begin{definition}
. The unit map is defined (as usual) by 
\[
I:{\bf C}\rightarrow A,\qquad i(\lambda )=\lambda 1;\quad 1\in A.
\]
\end{definition}

In a commutative diagrammatic representation the associativity of a ternary
algebra $A$ can be represented in the following way

\[
\begin{array}{c}
\,A\otimes {\bf C}\otimes {\bf C}\ ^{\underrightarrow{\,id\otimes \iota
\otimes i\,}}A\otimes \,\,A\otimes A \\ 
\,\simeq \downarrow
\,\,\,\,\,\,\,\,\,\,\,\,\,\,\,\,\,\,\,\,\,\,\,\,\,\,\,\,\,\,\,\,\downarrow _m
\\ 
A\,\,\,\,\,\,\,\,\,_{\overrightarrow{\,\,\,\,\,\,\,\,\,\,\,\,\,\,id\,\,\,\,}%
}\,\,\,\,\,\,\,\,A
\end{array}
\,\,\,\,\,\,\,\,\,\,\,\,,\,\,\,\,\,\,\,\,\,\,\,\,\,\,\,\,\,
\begin{array}{c}
\,\,{\bf C}\otimes A\otimes {\bf C}^{\underrightarrow{\,\iota \otimes
id\otimes i}}\,\,\,\,A\otimes A\otimes A \\ 
\,_{\simeq }\downarrow
\,\,\,\,\,\,\,\,\,\,\,\,\,\,\,\,\,\,\,\,\,\,\,\,\,\,\,\,\,\,\,\,\downarrow _m
\\ 
A\,\,\,\,\,\,\,\,_{\overrightarrow{\,\,\,\,\,\,\,id\,\,\,\,\,\,\,\,\,\,\,\,\,%
}}\,\,\,\,\,\,\,A
\end{array}
\]
\[
\begin{array}{c}
{\bf C}\otimes {\bf C}\otimes A\,\,\,^{\underrightarrow{i\otimes i\otimes id}%
}\,\,\,\,\,\,\,A\otimes A\,\otimes A\,\,\,\,\,\,\,\,\, \\ 
\simeq \downarrow
\,\,\,\,\,\,\,\,\,\,\,\,\,\,\,\,\,\,\,\,\,\,\,\,\,\,\,\,\,\,\,\,\,\,\,\,\,\,%
\downarrow _m \\ 
A\,\,\,\,\,\,\,\,\,\,\,_{\overrightarrow{\,\,\,\,\,\,\,\,\,\,\,id\,\,\,\,\,%
\,\,\,\,\,\,\,\,\,}\,\,\,\,\,\,\,\,\,\,\,}A
\end{array}
\]

On the other hand, the associativity of the multiplication is characterized
by

\[
m(m\otimes id\otimes id)=m(id\otimes m\otimes id)=m(id\otimes id\otimes m).
\]

\begin{definition}
. The Nambu product (a 3-commutator) is defined by $\pi =\pi ^{+}-\pi ^{-}$
, where 
\[
\pi ^{+}:A^{\otimes 3}\rightarrow A^{\otimes 3},\quad 
\]
\begin{equation}
\pi ^{+}(a,b,c)=a\otimes b\otimes c+b\otimes c\otimes a+c\otimes a\otimes b;
\end{equation}
\[
\pi ^{-}:A^{\otimes 3}\rightarrow A^{\otimes 3},\quad 
\]
\[
\pi ^{-}(a,b,c)=c\otimes b\otimes a+a\otimes c\otimes b+b\otimes a\otimes c.
\]
$\pi ^{+}(\pi ^{-})$ is the sum over all terms with even (odd) permutation
of $a,b$ and $c\in A.$
\end{definition}

Using this definition of Nambu product, a generalization of the concept of
abelian algebra ( an abelian 3-algebra) is given by the square representing
3-commutativity.

\[
\begin{array}{c}
\,\,\,\,\,\,\,\,A^{\otimes 3\,\,}\,\,\,\,\,\,\,^{\underrightarrow{%
\,\,\,\,\,\,\,\,\pi ^{+}\,\,\,\,\,\,}}\,\,\,\,\,\,\,A^{\otimes 3} \\ 
_{\pi ^{-}}\downarrow
\,\,\,\,\,\,\,\,\,\,\,\,\,\,\,\,\,\,\,\,\,\,\,\,\,\,\,\,\,\,\,\,\downarrow _m
\\ 
A^{\otimes 3}\,\,\,\,\,\,_{\overrightarrow{\,\,\,\,\,\,\,\,\,\,m\,\,\,\,\,\,%
\,\,\,}}\,\,\,\,\,\,\,\,A
\end{array}
\]

Notice that Definitions 1-4 can immediately be generalized for the $n$-ary
case. Besides, for the particular case $n=2$, we have 
\begin{eqnarray*}
m\circ \pi ^{+}(a\otimes b) &=&ab, \\
m\circ \pi ^{-}(a\otimes b) &=&ba.
\end{eqnarray*}
In this (binary) situation, $\pi ^{+}$ plays the role of the identity map ($%
id$), whilst $\pi ^{-}$ corresponds to the flip operator ($\tau $).

\begin{example}
{\sc .} {\em ( Abelian 3-algebra.) } Consider the set of functions from
Example 2, such that now 
\[
m(a\otimes b\otimes c){({\bf x)}}=a({\bf x})\,\,b({\bf x})\,\,c({\bf x}).
\]
In this case, the 3-commutator $\pi =0$.
\end{example}

\begin{example}
{\sc . }{\em (Noncommutative 3- algebra.)} Consider $A$ as given in
Example1. Then we can introduce the Nambu ternary bracket $[\cdot ,\cdot
,\cdot ]=m\circ \pi $ by the linear operator 
\[
\left[ a,b,c\right] =abc+bca+cab-cba-acb-bac,
\]
which satisfies the properties: \\ 
\end{example}

{\em (alternation law)}\\
\begin{equation}
\lbrack a,b,c]=[b,c,a]=[c,a,b]=-[a,c,b]=-[c,b,a]=-[b,a,c],  \label{alt}
\end{equation}

{\em (derivation law) } 
\begin{equation}
\lbrack a,b,cd]=c[a,b,d]+[a,b,c]d,  \label{der}
\end{equation}

{\em ( generalized Jacobi identity)} 
\begin{equation}
\lbrack g,h[a,b,c]]=[[g,h,a],b,c]+[a,[g,h,b],c]+[a,b,[g,h,c]].  \label{jac}
\end{equation}
Such a generalized Jacobi identity has been analyzed by several authors\cite
{rugg, flato4, tacht1, ad1}, in different contexts, and it was called {\it %
fundamental identity} in Ref.\cite{tacht1}.

\begin{example}
. (Commutative 3-algebra.) Consider $A$ as given by Example 2(in this case $m
$ is a nonassociative product). Then we can introduce the Nambu bracket $%
\left\{ \cdot ,\cdot ,\cdot \right\} =m\circ \pi $ by $\left\{ a,b,c\right\}
=\varepsilon ^{ijk}\,\partial _ia\,\,\partial _jb\,\,\partial _kc$. A basis
for this (classical) Nambu bracket is then $x_{1,\,\,}x_2,x_{3,\;\;\;}$such
that $\left\{ x_1,x_2,x_2\right\} =1.$
\end{example}

We use the definition of 3-algebra, given in terms of commutative diagrams,
to explore the structure of dual coalgebra, and so to introduce a
generalization of Hopf algebra \cite{ad2, madore}. To do this, we proceed as
usually is done with the concept of coalgebra: a 3-coproduct is defined by
inverting the arrows in the definition of the 3-associative algebra.
Therefore, we obtain the definition of 3-comultiplication $\Delta $ and
3-counit $\epsilon $, 
\[
\Delta :A\rightarrow A^{\otimes 3}, 
\]
\[
\epsilon :A\rightarrow {\bf C,} 
\]
such that the following diagrams commute

\[
\begin{array}{c}
\,A\otimes {\bf C}\otimes {\bf C}\ ^{\underleftarrow{id\otimes \epsilon
\otimes \epsilon }}A^{\otimes 3} \\ 
\,_{\simeq }\uparrow
\,\,\,\,\,\,\,\,\,\,\,\,\,\,\,\,\,\,\,\,\,\,\,\,\,\,\,\,\,\,\,\,\uparrow
_\Delta  \\ 
A\,\,\,\,\,\,_{\overleftarrow{\,\,\,\,\,\,\,\,\,id\,\,\,\,\,\,\,\,\,\,\,\,}%
}\,\,\,\,\,\,\,\,A
\end{array}
\,\,\,\,\,\,\,\,\,\,\,\,\,\,\,\,\,\,\,\,\,\,\,\,\,\,\,\,\,
\begin{array}{c}
\,\,{\bf C}\otimes A\otimes {\bf C}^{\underleftarrow{\epsilon \otimes
id\otimes \epsilon }}\,\,\,A^{\otimes 3} \\ 
_{\simeq }\uparrow
\,\,\,\,\,\,\,\,\,\,\,\,\,\,\,\,\,\,\,\,\,\,\,\,\,\,\,\,\,\,\,\,\uparrow
_\Delta  \\ 
A\,\,\,\,_{\,\overleftarrow{\,\,\,\,\,\,\,\,\,\,\,\,\,\,\,id\,\,\,\,\,\,\,\,%
\,\,\,\,\,\,}}\,\,\,\,\,\,\,\,A
\end{array}
\]
\[
\begin{array}{c}
{\bf C}\otimes {\bf C}\otimes A\,\,^{\underleftarrow{\epsilon \otimes
\epsilon \otimes id}}\,\,\,\,\,\,A^{\otimes 3}\,\,\,\,\,\,\,\,\, \\ 
\,_{\simeq }\uparrow
\,\,\,\,\,\,\,\,\,\,\,\,\,\,\,\,\,\,\,\,\,\,\,\,\,\,\,\,\,\,\,\,\,\,\,\,\,\,%
\uparrow _\Delta  \\ 
A\,\,\,\,\,_{\overleftarrow{\,\,\,\,\,\,\,\,\,\,\,\,\,\,\,id\,\,\,\,\,\,\,\,%
\,\,\,\,}}\,\,\,\,\,\,\,\,\,\,A
\end{array}
\]

The 3-coassociativity is defined by

\[
(\Delta \otimes id\otimes id)\Delta =(id\otimes \Delta \otimes id)\Delta
=(id\otimes id\otimes \Delta )\Delta ,
\]

and 3-cocommutativity is expressed by square

\[
\begin{array}{c}
\,\,\,\,\,\,\,\,\,A^{\otimes 3\,\,}\,\,\,\,\,\,^{\underleftarrow{%
\,\,\,\,\,\,\,\,\,\,\,\,\,\,\pi ^{+}\,\,\,\,\,}}\,\,\,\
\,\,\,\,\,\,\,\,A^{\otimes 3} \\ 
_{\pi ^{-}}\uparrow
\,\,\,\,\,\,\,\,\,\,\,\,\,\,\,\,\,\,\,\,\,\,\,\,\,\,\,\,\,\,\,\,\,\,\,\,%
\uparrow _\Delta  \\ 
A^{\otimes 3}\,\,\,_{\,\,\overleftarrow{\,\,\,\,\,\,\,\,\,\,\,\,\,\Delta
\,\,\,\,\,\,\,\,\,}}\,\,\,\,\,A
\end{array}
\]

A 3-algebra and a 3-coalgebra give rise to a generalization of bialgebras
(that is, a 3-bialgebra). In order to define a Hopf 3-algebra, a
generalization of antipode should be introduced. A natural 3-antipode $S$
can be defined as follows, 
\begin{eqnarray}
m\circ (S\otimes id\otimes id)\circ \Delta  &=&i\circ \epsilon ,
\label{anti1} \\
m\circ (id\otimes S\otimes id)\circ \Delta  &=&i\circ \epsilon ,
\label{anti2} \\
m\circ (id\otimes id\otimes S)\circ \Delta  &=&i\circ \epsilon \,.
\label{anti3}
\end{eqnarray}
For 4-bialgebras, a 4-antipode can be introduced by 
\begin{eqnarray}
m\circ (S\otimes id\otimes S\otimes id)\circ \Delta  &=&i\circ \epsilon ,
\label{an1} \\
m\circ (id\otimes S\otimes S\otimes id)\circ \Delta  &=&i\circ \epsilon ,
\label{an2} \\
m\circ (S\otimes id\otimes id\otimes S)\circ \Delta  &=&i\circ \epsilon ,
\label{an3} \\
m\circ (id\otimes S\otimes id\otimes S)\circ \Delta  &=&i\circ \epsilon .
\label{an4}
\end{eqnarray}
All such relations are supposed to be satisfied simultaneously

\begin{example}
{\sc .} {\em ( 3-bialgebra from Nambu-Lie algebra.)} Considering $A$ in
Example 4, a 3-bialgebra can be derived, if we introduce a 3-coproduct by
\end{example}

\[
\Delta a=a\otimes a\otimes a\equiv a^{\otimes 3}. 
\]
Indeed, in this case $\Delta $ respects the algebraic relation, since

\[
\lbrack \Delta a,\Delta b,\Delta c]=\Delta [a,b,c]. 
\]
Notice that if we try to introduce a 3-coproduct by 
\begin{equation}
\Delta a=a\otimes 1\otimes 1+1\otimes a\otimes 1+1\otimes 1\otimes a,
\end{equation}
the algebra structure can not be respected, because $\Delta [a,b,c]\ne
[\Delta a,\Delta b,\Delta c]$. This result shows us that we can obtain a
bialgebra structure attached to some (if any) universal enveloping 3-algebra
of a Nambu-Lie algebra, but a Hopf 3-algebra can not be trivially introduced
in this case.

\begin{example}
{\sc .} {({\em $SL(n,{\bf C})$)}} Consider the group $G=SL(n,{\bf C),}$
where an element $x=(a_j^i)\in G$ has unit determinant, $det\,\,x=1$. It is
well known that the algebra generated by the functions $a_j^i(x)$, with
2-coproduct defined by $\Delta {a^i}_j=\sum_k{a^i}_k\otimes {a^k}_j$ is a
Hopf algebra (see, for example, \cite{madore}), with counit given by 
\begin{equation}
\epsilon ({a^i}_j)=\delta {^i}_j  \label{counit11}
\end{equation}
and the antipode given by inverse matrix 
\begin{equation}
S({a^i}_j)={(a^{-1})^i}_j.  \label{ant6}
\end{equation}
In order to study a generalization of this Hopf algebra, we can consider two
cases, 3- and 4- algebras. First, a 3-coproduct $\Delta $ can be defined as
following 
\begin{equation}
\Delta a_j^i(x,y,z)=a_j^i(x,y,z)=\sum_{k,l}a_k^i(x)a_l^k(y)a_j^l(z)
\end{equation}
\[
=\sum_{k,l}a_k^i\otimes a_l^k\otimes a_j^l(x,y,z),
\]
or 
\begin{equation}
\Delta a_j^i=\sum_{k,l}a_k^i\otimes a_l^k\otimes a_j^l.
\end{equation}
Therefore, a 3-bialgebra can be derived, using the usual counit, Eq.(\ref
{counit11}). It is an easy matter to see that the usual antipode, Eq.(\ref
{ant6}), can not be used to define a satisfactory 3-antipode as given by
Eqs.(\ref{anti1})--(\ref{anti3}).

For 4-product, however, using 
\[
\Delta a_j^i=\sum_{k,l,m}a_k^i\otimes a_l^k\otimes a_m^l\otimes a_j^m,
\]
together with Eqs.(\ref{an1})--(\ref{an4}), where $S$ is being given by Eq.(%
\ref{ant6}), we get a Hopf 4-algebra.
\end{example}

In short, we have presented here a generalization of the concept of
associative and commutative algebra. Exploring the notion of duality, and
following in parallel with the usual (binary) approach, the concept of
n-bialgebra could be introduced. In particular, we have presented examples
of n-bialgebras, as that one derived from the Nambu-Lie algebra and that
associated with the $SL(n,C)$ group. It has also been indicated how to
introduce the concept of 3- and 4-antipode, in order to obtain a
generalization of Hopf algebra. As an example, a Hopf 4-algebra attached to $%
SL(n,C)$ was derived. It should be interesting to study the connection of a
3-Hopf algebra with a universal enveloping algebra of a 3-noncommutative
algebra. These aspects will be studied in more details elsewhere.

\subparagraph{
\[
\]
}

{\bf Acknowledgments}

We are grateful to A. Matos Neto and J. David M. Vianna for helpful
discussions. One of us (A.E.S) thanks to CNPq ( a Brazilian Agency for
Research) for financial support.

\end{document}